\documentclass{article}

\usepackage{amsfonts}
\usepackage{float}
\usepackage{amsmath}
\usepackage{amssymb}
\usepackage{amsthm}
\usepackage{graphicx}
\usepackage{multirow}
\usepackage{color}
\usepackage{leqno}
\usepackage{ulem}
\usepackage{setspace}
\usepackage{caption}
\usepackage{subcaption}
\usepackage{rotating}
\usepackage{threeparttable}
\usepackage{url}

\newcommand{\mbf}[1]{\mbox{\boldmath $#1$}}
\newcommand{\ba}{{\mbf \beta}}

\def\bfX{{\mathbf X}}

\def\bfY{{\mathbf Y}}

{\catcode `\@=11 \global\let\AddToReset=\@addtoreset}
\AddToReset{equation}{section}

\AddToReset{Theorem}{section}

\theoremstyle{remark}

\def\bc{\begin{center}}
\def\bd{\begin{description}}
\def\be{\begin{enumerate}}
\def\ec{\end{center}}
\def\ed{\end{description}}
\def\edt{\end{document}}
\def\ee{\end{enumerate}}
\def\ben{\begin{equation}}
\def\benn{\begin{equation*}}
\def\een{\end{equation}}
\def\eenn{\end{equation*}}
\def\benr{\begin{eqnarray}}
\def\eenr{\end{eqnarray}}
\def\benrr{\begin{eqnarray*}}
\def\eenrr{\end{eqnarray*}}

\def\b{\beta}

\def\del{\delta}
\def\edt{\end{document}}

\def\lel{\label}

\def\r{\ref}

\def\vep{\varepsilon}

\def\vs{\vskip}
\def\wh{\widehat}

\def\wt{\widetilde}

\def\R{{\mathbb R}}

\def\bc{\begin{center}}
\def\ec{\end{center}}

\begin{document}

\bc
{\large A Fast Algorithm for Solving Henderson's Mixed Model Equation}\\[.2cm]
\large{Jiwoong Kim\\ Michigan State University}
\ec
\vs .75cm

\begin{abstract}
\noindent
This article investigates a fast and stable method to solve Henderson's mixed model equation. The proposed algorithm is stable in that it avoids inverting a matrix of a large dimension and hence is free from the curse of dimensionality. This tactic is enabled through row operations performed on the design matrix.
\noindent\\
\textit{Keywords}: Henderson's mixed model equation; inverting matrix; row operations
\end{abstract}

\section{Introduction}\lel{Sec:Intro}
A linear mixed model is a model which contains fixed effects and unobservable random effects. Consider the linear mixed model
\benr\lel{eq:Model1}
Y_{ij} &=& \mbf{x}_{ij}'\ba+v_{i} + \vep_{ij},\quad\quad\quad\quad i=1,2,...,n;\,\,\,\,j=1,2,...,m,\nonumber
\eenr
where $\mbf{x}_{ij}=(x_{ij}^{1},...,x_{ij}^{p})'\in \R^{p}$ are non-random design variables, and $\ba=(\beta_{1},...,\beta_{p})'\in \R^{p}$ is a parameter vector of interest. $v_{i}$'s are unobservable random effects, and $\vep_{ij}$'s are errors which are independent from the random effects. Define
\benrr
&&\bfY_i =\left[
         \begin{array}{c}
           Y_{i1} \\
           Y_{i2} \\
           \vdots \\
           Y_{im} \\
         \end{array}
       \right]_{m\times 1},\quad \bfX_{i} = \left[
                              \begin{array}{cccc}
                                x_{i1}^{1} & x_{i1}^{2} & \cdots & x_{i1}^{p} \\
                                x_{i2}^{1} & x_{i2}^{2} & \cdots & x_{i2}^{p} \\
                                \vdots & \vdots & \ddots & \vdots \\
                                x_{im}^{1} & x_{im}^{2} & \cdots & x_{im}^{p} \\
                              \end{array}
                            \right]_{m\times p},\quad \mbf{\vep}_{i}= \left[
         \begin{array}{c}
           \vep_{i1} \\
           \vep_{i2} \\
           \vdots \\
           \vep_{im} \\
         \end{array}
       \right]_{m\times 1},\quad \mbf{1}_{m}=\left[
         \begin{array}{c}
           1 \\
           1 \\
           \vdots \\
           1 \\
         \end{array}
       \right]_{m\times 1},\\
&&\bfY=\left[
         \begin{array}{c}
           \bfY_{1} \\
           \bfY_{2} \\
           \vdots \\
           \bfY_{n} \\
         \end{array}
       \right]_{nm\times 1}, \quad \bfX = \left[
         \begin{array}{c}
           \bfX_{1} \\
           \bfX_{2} \\
           \vdots \\
           \bfX_{n} \\
         \end{array}
       \right]_{nm\times p},\quad
       \mbf{Z} = \left[
                              \begin{array}{cccc}
                                \mbf{1}_{m} & \mbf{0}_{m} & \cdots & \mbf{0}_{m} \\
                                \mbf{0}_{m} & \mbf{1}_{m} & \cdots & \mbf{0}_{m} \\
                                \vdots & \vdots & \ddots & \vdots \\
                                \mbf{0}_{m} & \mbf{0}_{m} & \cdots & \mbf{1}_{m} \\
                              \end{array}
                            \right]_{mn\times n}, 
\eenrr
\benrr
&&\mbf{v}=\left[
         \begin{array}{c}
           v_{1} \\
           v_{2} \\
           \vdots \\
           v_{n} \\
         \end{array}
       \right]_{n\times 1}, \quad \mbf{\vep} = \left[
         \begin{array}{c}
           \mbf{\vep}_{1} \\
           \mbf{\vep}_{2} \\
           \vdots \\
           \mbf{\vep}_{n} \\
         \end{array}
       \right]_{nm\times 1}.
\eenrr
Then the model (\r{eq:Model1}) can be expressed as
\benn
\bfY = \bfX\ba+\mbf{Z}\mbf{v}+\mbf{\vep}.
\eenn
Various authors proposed the best linear unbiased estimates of fixed effects and the best linear unbiased predictions of random effects: see, e.g., [1], [2], and [3]. Assuming that 
\benn
\mbf{v}\sim N(\mbf{0}_{n\times 1}, \mbf{I}_{n\times n}),\qquad \mbf{\vep}\sim N(\mbf{0}_{nm\times 1}, \mbf{I}_{nm\times nm}),
\eenn
i.e., $\phi=\lambda = 1$ for the simplicity, and maximizing the joint density of $\bfY$ and $\mbf{v}$ yield Henderson's mixed model equations 
\benn
\mbf{A}\mbf{\del} = \mbf{c}
\eenn
where
\benn
\mbf{A} = \left[
  \begin{array}{cc}
    \bfX^{T}\bfX & \bfX^{T}\mbf{Z} \\
    \mbf{Z}^{T}\bfX & \mbf{Z}^{T}\mbf{Z}+\mbf{I} \\
  \end{array}
\right],\quad \mbf{\del} = \left[
  \begin{array}{c}
    \wh{\mbf{\b}} \\
    \wh{\mbf{v}} \\
  \end{array}
\right],\quad \mbf{c} = \left[
\begin{array}{c}
    \bfX^{T}\bfY \\
    \mbf{Z}^{T}\bfY \\
  \end{array}
\right].
\eenn
The solutions to the equations are the best linear unbiased estimates and predictors for $\mbf{\b}$ and $\mbf{v}$, respectively. This article proposes the fast and stable method for the solutions; we, however, consider only normal random effect and error. The proposed algorithm can be applied to other cases in the similar manner.

\section{Algorithm: transformation through row operations}
Define a $(p+n)\times (p+n+1)$ new matrix
\benrr
\mbf{D} &=& \left[
  \begin{array}{c|c}
    \mbf{A} & \mbf{c} \\
  \end{array}
\right]\\
        &=&  \left[
  \begin{array}{cc|c}
    \bfX^{T}\bfX & \bfX^{T}\mbf{Z} & \bfX^{T}\mbf{Y}\\
    \mbf{Z}^{T}\bfX &\mbf{Z}^{T}\mbf{Z}+\mbf{I} & \mbf{Z}^{T}\mbf{Y}\\
  \end{array}
\right],
\eenrr
where dimensions of block matrices are $p\times p$, $p\times n$, $p\times 1$, $n\times p$, $n\times n$, and $n\times 1$, respectively.
Note that $\mbf{Z}^{T}\mbf{Z}+\mbf{I}$ is a $n\times n$ diagonal matrix whose diagonal entry is $(m+1)$. Also we have
\benrr
\mbf{Z}^{T}\bfX &=&  \left[
         \begin{array}{c}
           \mbf{1}_{m}^{T}\bfX_{1} \\
           \mbf{1}_{m}^{T}\bfX_{2} \\
           \vdots \\
           \mbf{1}_{m}^{T}\bfX_{n} \\
         \end{array}
       \right]_{n\times p} = \left[
         \begin{array}{c}
           \sum_{j=1}^{m} \mbf{x}_{1j}^{T} \\
           \sum_{j=1}^{m} \mbf{x}_{2j}^{T} \\
           \vdots \\
           \sum_{j=1}^{m} \mbf{x}_{nj}^{T} \\
         \end{array}
       \right]_{n\times p}.
\eenrr
These two facts will be rigorously exploited in the proposed algorithm: the proposed algorithm does not require $\mbf{Z}$ which hinders a fast computation when $n$ is relatively large. Through the row operations, the proposed algorithm transforms $\mbf{D}$ into
\benrr
\widetilde{\mbf{D}} &=& \left[
  \begin{array}{cc|c}
    \wt{\bfX} & \mbf{0} & \widetilde{\mbf{c}}_{1}\\
    \mbf{Z}^{T}\bfX &\mbf{Z}^{T}\mbf{Z}+\mbf{I} & \mbf{Z}^{T}\mbf{Y}\\
  \end{array}
\right]
\eenrr
so that
\benn
\wh{\mbf{\b}} = (\wt{\bfX})^{-1}\widetilde{\mbf{c}}_{1}
\eenn
where $\wt{\bfX}$ is a $p\times p$ nonsingular matrix. The row operations can further be performed so that the inverse of the matrix is not necessary when $p$ is large. The computation of the inverse of $\wt{\bfX}$ is reasonably fast till $p=5,000$. Beyond $p=5,000$, the further row operations are recommended. Then,
\benn
\wh{\mbf{v}} = (m+1)^{-1}(\mbf{Z}^{T}\mbf{Y} - \mbf{Z}^{T}\bfX \wh{\mbf{\b}} ).
\eenn
Let $\mbf{D}^{(k)}$ denote the matrix $\mbf{D}$ at the $k$th stage of the row operations. Next, we shall partition it into six blocks:
\benrr
\mbf{D}^{(k)}     &=&  \left[
  \begin{array}{cc|c}
    \mbf{D}_{11}^{(k)} & \mbf{D}_{12}^{(k)} & \mbf{D}_{13}^{(k)}\\
    \mbf{D}_{21}^{(k)} & \mbf{D}_{22}^{(k)} & \mbf{D}_{23}^{(k)}\\
  \end{array}
\right].
\eenrr
Observe that $\mbf{D}_{21}^{(k)}$, $\mbf{D}_{22}^{(k)}$, and $\mbf{D}_{23}^{(k)}$ will remain intact, which implies that they are equal to
$\mbf{Z}^{T}\bfX$, $\mbf{Z}^{T}\mbf{Z}+\mbf{I}$, and $\mbf{Z}^{T}\mbf{Y}$ through whole stages, respectively. Let $d_{22}$ denote the diagonal entry of $\mbf{D}_{22}^{(k)}$. Also let $\mbf{D}_{ij}^{(k)}[l,:]$ and $\mbf{D}_{ij}^{(k)}[l, m]$ denote the $l$th row vector and the $(l,m)$th entry of the matrix $\mbf{D}_{ij}^{(k)}$ for $i=1,2$ and $j=1,2,3$, respectively. As will be shown later, actual row operations are performed on $\mbf{D}_{11}^{(k)}$ and $\mbf{D}_{13}^{(k)}$ only.

The following is the summary of the proposed algorithm for transforming $\mbf{D}$ into $\wt{\mbf{D}}$, that is, transforming $\mbf{D}_{12}$ into $\mbf{D}_{12}^{(n)}=\mbf{0}_{p\times n}$.
\begin{table}[h!]
\centering
\begin{tabular}{l}
  \hline
  \textbf{The proposed algorithm: } \\
  \hline
  for $k=1$ to $n$\\
    \\
    \quad for $h=1$ to $p$\\
    \\
    \quad\quad $c_{kh} = \mbf{D}_{12}^{(k)}[p-h,n-k]/d22$\\
    \quad\quad $\mbf{D}_{11}^{(k)}[p-h,:] = \mbf{D}_{11}^{(k)}[p-h,:] - c_{kh} \mbf{D}_{21}^{(k)}[n-k,:]$\\
    \quad\quad $\mbf{D}_{13}^{(k)}[p-h,:] = \mbf{D}_{13}^{(k)}[p-h,:] - c_{kh} \mbf{D}_{23}^{(k)}[n-k,:]$\\
    \\
    \quad end for\\
    \\
  \quad Update $\mbf{D}^{(k)}$ to $\mbf{D}^{(k+1)}$\\
  end for\\
  \hline
\end{tabular}
\end{table}

\section{Computational time}
Table \r{table:1cpu} reports computational times of the proposed algorithm when $n$ and $p$ vary with $m$ being fixed at 10. The proposed algorithm is iterated 10 times, and the average cpu time of 10 iterations is reported. The cpu used in this simulation is Intel Core i5-3570 3.40 GHz.
\begin{table}[h!]
\centering
\begin{tabular}{c  c c c c }
\hline
      & $p=$10   &  20 & 100  & 200  \\
 \hline
 $n = $1,000  & 0.004 & 0.005 & 0.078 & 0.297\\
 \hline
 2,000  & 0.005 & 0.010 &0.161 &0.645\\
 \hline
 5,000 & 0.008 & 0.026  &0.392 &1.496\\
 \hline
 10,000  &0.023 &0.057  &0.794 &3.221\\
 \hline
\end{tabular}
\caption{cpu times when $n$ and $p$ vary.}\lel{table:1cpu}
\end{table}
As reported in the table, we can see that computational time is $O(n)$. When a dimension of the design matrix $\bfX$ is $10^{5}\times 200$ ($n=10^{4}, m=10, p=200$), it takes only 3.221 cpu seconds.

\edt